\RequirePackage{fix-cm}
\documentclass[english]{IEEEtran}
\usepackage[T1]{fontenc}
\usepackage[latin9]{inputenc}
\usepackage{float}
\usepackage{amsthm}
\usepackage{amsmath}
\usepackage{amssymb}
\usepackage{graphicx}
\PassOptionsToPackage{version=3}{mhchem}
\usepackage{mhchem}
\usepackage{babel}

\usepackage[crop=off]{auto-pst-pdf}
\ifpdf
  \usepackage{forest}
\else
  \usepackage{pst-barcode}
\fi

\makeatletter

\floatstyle{ruled}
\newfloat{algorithm}{tbp}{loa}
\providecommand{\algorithmname}{Algorithm}
\floatname{algorithm}{\protect\algorithmname}

\theoremstyle{plain}
\newtheorem{thm}{\protect\theoremname}
\theoremstyle{remark}
\newtheorem{rem}[thm]{\protect\remarkname}

\providecommand{\remarkname}{Remark}
\providecommand{\theoremname}{Theorem}

\begin{document}

\title{Joint Interference Alignment and Bi-Directional Scheduling for MIMO
Two-Way Multi-Link Networks}

\author{A. M. Fouladgar, O. Simeone
\thanks{A. M. Fouladgar and O. Simeone are with the CWCSPR, New Jersey Institute
of Technology, Newark, NJ 07102 USA (e-mail: \{af82,osvaldo.simeone\}@njit.edu).%
}, O. Sahin
\thanks{O. Sahin is with InterDigital Inc., Melville, New York, 11747, USA
(email: Onur.Sahin@interdigital.com).%
}, P. Popovski%
\thanks{P. Popovski is with the Department of Electronic Systems Aalborg University,
Denmark (e-mail: petarp@es.aau.dk).%
} and S. Shamai (Shitz)%
\thanks{S. Shamai (Shitz) is with the Department of Electrical Engineering,
Technion, Haifa, 32000, Israel (email: sshlomo@ee.technion.ac.il).%
}}
\maketitle
\begin{abstract}
By means of the emerging technique of dynamic Time Division Duplex
(TDD), the switching point between uplink and downlink transmissions
can be optimized across a multi-cell system in order to reduce the
impact of inter-cell interference. It has been recently recognized
that optimizing also the order in which uplink and downlink transmissions,
or more generally the two directions of a two-way link, are scheduled
can lead to significant benefits in terms of interference reduction.
In this work, the optimization of bi-directional scheduling is investigated
in conjunction with the design of linear precoding and equalization
for a general multi-link MIMO two-way system. A simple algorithm is
proposed that performs the joint optimization of the ordering of the
transmissions in the two directions of the two-way links and of the
linear transceivers, with the aim of minimizing the interference leakage
power. Numerical results demonstrate the effectiveness of the proposed
strategy.

\end{abstract}

\section{Introduction}

\begin{figure}
\centering\includegraphics[clip,scale=0.45]{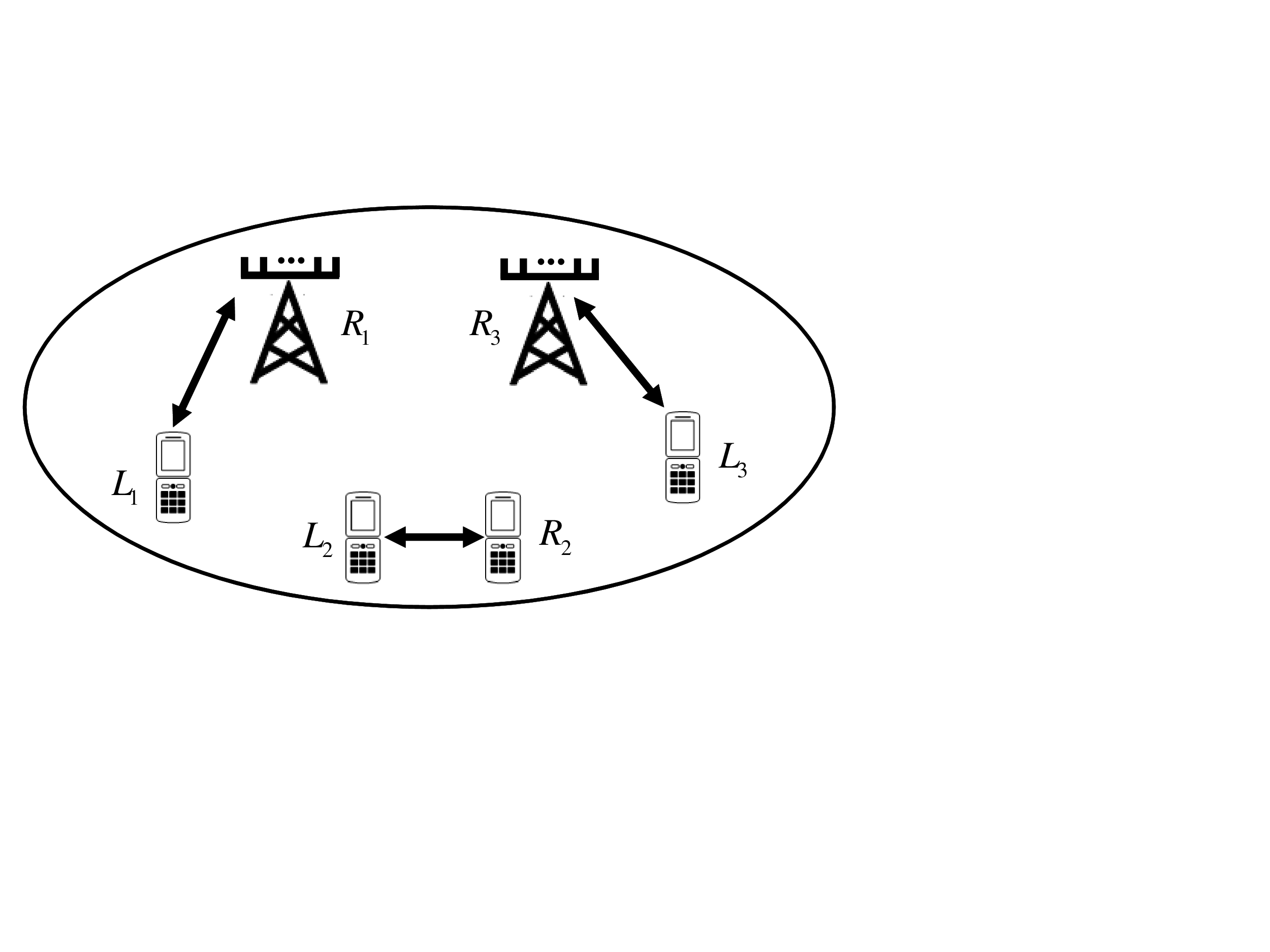}\protect\protect\protect\caption{\label{fig:1}A network of two-way MIMO communication links. Examples
include uplink-downlink cellular links and two-way D2D channels.}
\end{figure}

\textit{\emph{Two-way communication is one of the most common modes
of operation for wireless links, particularly for cellular and Device-to-Device
(D2D) systems (see Fig. \ref{fig:1}). The conventional approach to
the design of a system comprising multiple interfering two-way links
consists of two separate phases: at first, one fixes a transmission
direction independently for each link; and, then, the physical-layer
parameters, such as powers or beamforming vectors, are optimized in
a centralized or decentralized fashion so as to maximize some system-wide
performance criterion. This is, for instance, the standard approach
for cellular systems, in which the scheduling of uplink and/or downlink
transmissions is performed on a per-cell basis as a preliminary step.}}

\textit{\emph{The emerging technique of dynamic Time Division Duplex
(TDD) breaks with the conventional approach discussed above in that
the switching points between uplink and downlink transmissions in
a frame are optimized jointly across all cells based on the current
channel conditions}}\emph{ }\cite{Yu}\textendash \cite{Bamby}.\textit{\emph{
The key motivation for this paradigm change is the observation that
the selection of the duration of the uplink and downlink transmission
phases in a cell can have a significant impact on the interference
observed by the interfered cells. }}

\begin{figure}
\centering\includegraphics[clip,scale=0.4]{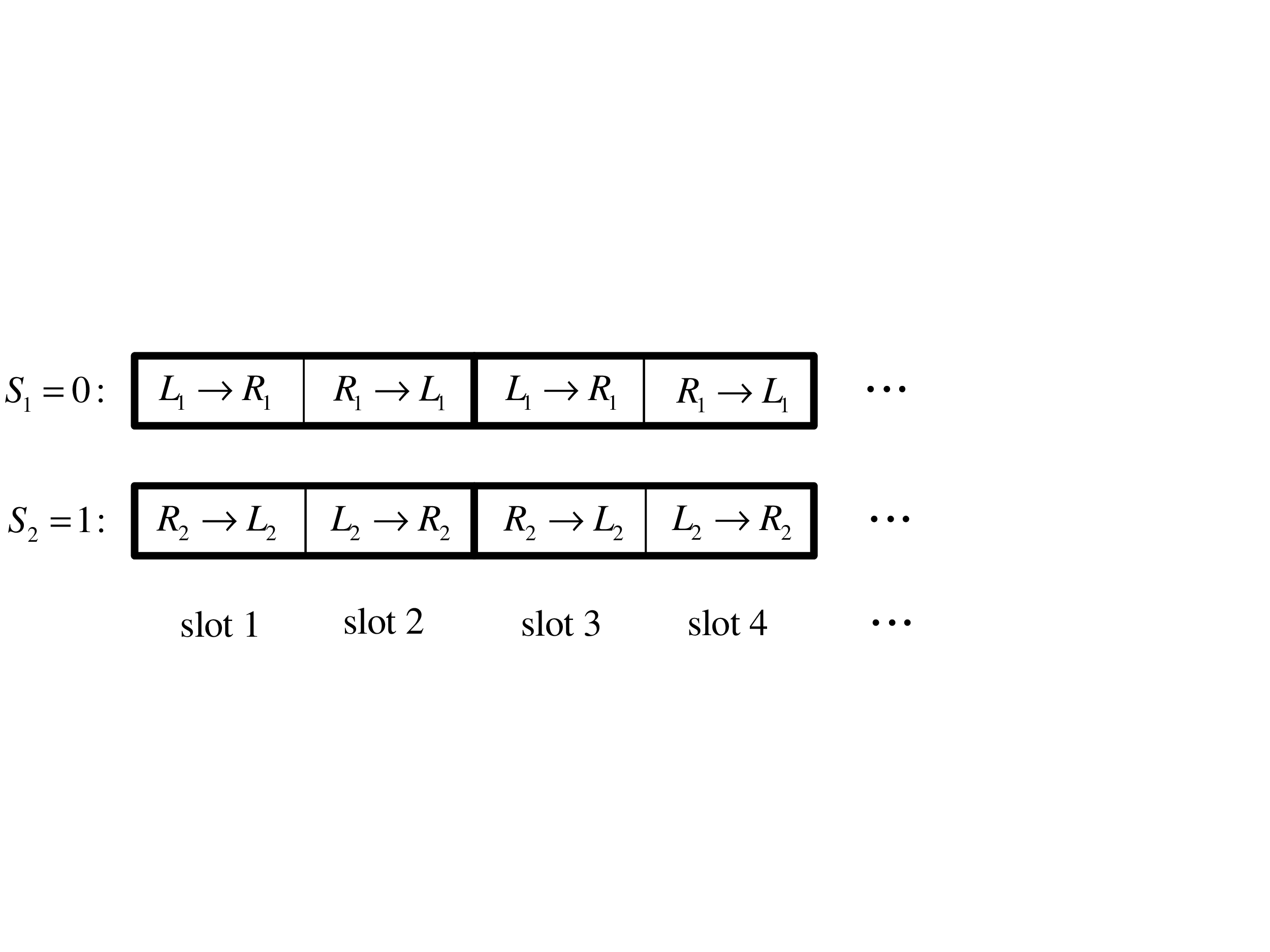}

\protect\protect\protect\caption{\label{fig:2}Illustration of the definition of interference spin
$S_{i}$ for each link $i$. A frame consists of two consecutive slots
as shown by bold lines.}
\end{figure}

\textit{\emph{With dynamic TDD, the ordering of the transmissions
in the two directions of a two-way link is fixed. However, the interference
configuration depends significantly on such scheduling decisions.
An illustration of bi-directional scheduling is provided in Fig. \ref{fig:2}
for two links: in each frame comprising two slots of fixed duration,
each link can operate in the left ($L$) to right ($R$) direction,
i.e., $L\rightarrow R$, in the first slot and then in the opposite
direction, $R\rightarrow L$, in the second slot, or it can schedule
first the $R\rightarrow L$ direction and then the $L\rightarrow R$
direction. As in \cite{Jimmy}, we refer to this binary choice for
each link as the }}\textit{interference spin}\textit{\emph{ of the
link.}}

\textit{\emph{Reference \cite{Jimmy} studies the optimization of
bi-directional scheduling in the presence of fixed switching times
for single-antenna links. In this work, instead, we investigate the
interplay of multi-antenna transmission and reception with bi-directional
scheduling for fixed switching times and a general MIMO two-way multi-link
system. }}This study is motivated by the observation that the choice
of the interference spins across the network defines the channel matrices
of the interfering channels on which the linear transceivers operate.
Therefore, the capability of multiple antennas to mitigate interference
depends on the scheduling decisions in a fundamental way. Algorithms
are proposed that perform either the separate or the joint optimization
of the interference spins and of the linear transceivers, with the
aim of minimizing the interference leakage power and hence obtaining
enhanced interference alignment solutions in the sense of \cite{Jafar3}.

\textit{Notation: }We use lower case fonts for scalars and uppercase
bold fonts for vectors and matrices. $\mathbf{I}_{d}$ represents
the $d\times d$ identity matrix and $\mathbf{0}_{M\times N}$ is
used to indicate the $M\times N$ zero matrix. $\mathrm{tr}\left(\mathbf{A}\right)$
denotes the trace of the matrix $\mathbf{A}$ and $\mathbf{A}^{H}$
is the conjugate transpose of matrix $\mathbf{A}$. $\nu_{\min}^{d}(\cdot)$
returns a truncated unitary matrix spanning the space associated with
the $d$ smallest eigenvalues of the argument matrix. $\mathbb{C}^{M\times N}$
represents the set of complex-valued $M\times N$ matrices and $\mathbb{U}^{M\times N}$
represents the set of truncated unitary matrices with $N$ orthonormal
columns. $\bar{X}$ is the complement of binary variable $X$. The
indicator function $1[\cdot]$ returns one if the argument is true
and it returns zero otherwise.

\section{System Model\label{sec:System-Model}}

We consider a wireless network consisting of $K$ MIMO two-way interfering
links. An illustration is shown in Fig. \ref{fig:1} for a scenario
with two cellular links and a D2D link. Each two-way link $k$ is
equipped with $N_{L,k}$ antennas at the left-hand node $L_{k}$ and
with $N_{R,k}$ antennas at the right-hand node $R_{k}$. Note that
the labeling of one end of each link as ``right'' or ``left''
is arbitrary. Each link operates using Time Division Duplex (TDD),
and two-way communication takes a frame consisting of two successive
slots: in the first slot, each $k$th link may operate either in the
direction $L_{k}\rightarrow R_{k}$, so that the $L_{k}$ node is
the transmitter and the $R_{k}$ node is the receiver, or, vice versa,
in the direction $R_{k}\rightarrow L_{k}$, and the successive slot
is used in the opposite direction. All links are synchronous and are
assumed to be always backlogged so that there is a continuous stream
of frames.

Based on the discussion above, each link $k$ operates in either direction
$L_{k}\rightarrow R_{k}$ or $R_{k}\rightarrow L_{k}$ in even slots
and in the opposite direction in odd slots. Following \cite{Jimmy},
we define the order in which the two directions are scheduled as the
\textit{interference spin} or, for short, \textit{spin} of a link.
Specifically, the $k$-th link is said to have a 0-spin if it operates
in the direction $L_{k}\rightarrow R_{k}$ in the odd slots and the
direction $R_{k}\rightarrow L_{k}$ in the even slots; otherwise,
when it operates in the direction $R_{k}\rightarrow L_{k}$ in the
odd slots and in the direction $L_{k}\rightarrow R_{k}$ in the even
slots, the $k$-th link is said to have a 1-spin. We refer to Fig.
\ref{fig:2} for an illustration. The spin of the $k$-th link is
denoted by $S_{k}\in\{0,1\}$. In order to simplify the notation,
in the following, we set $N_{L,k}=N_{L}$ and $N_{R,k}=N_{R}$ for
all $k\in{\cal K}$, although the generalization is straightforward.
\begin{figure}
\centering\includegraphics[clip,scale=0.5]{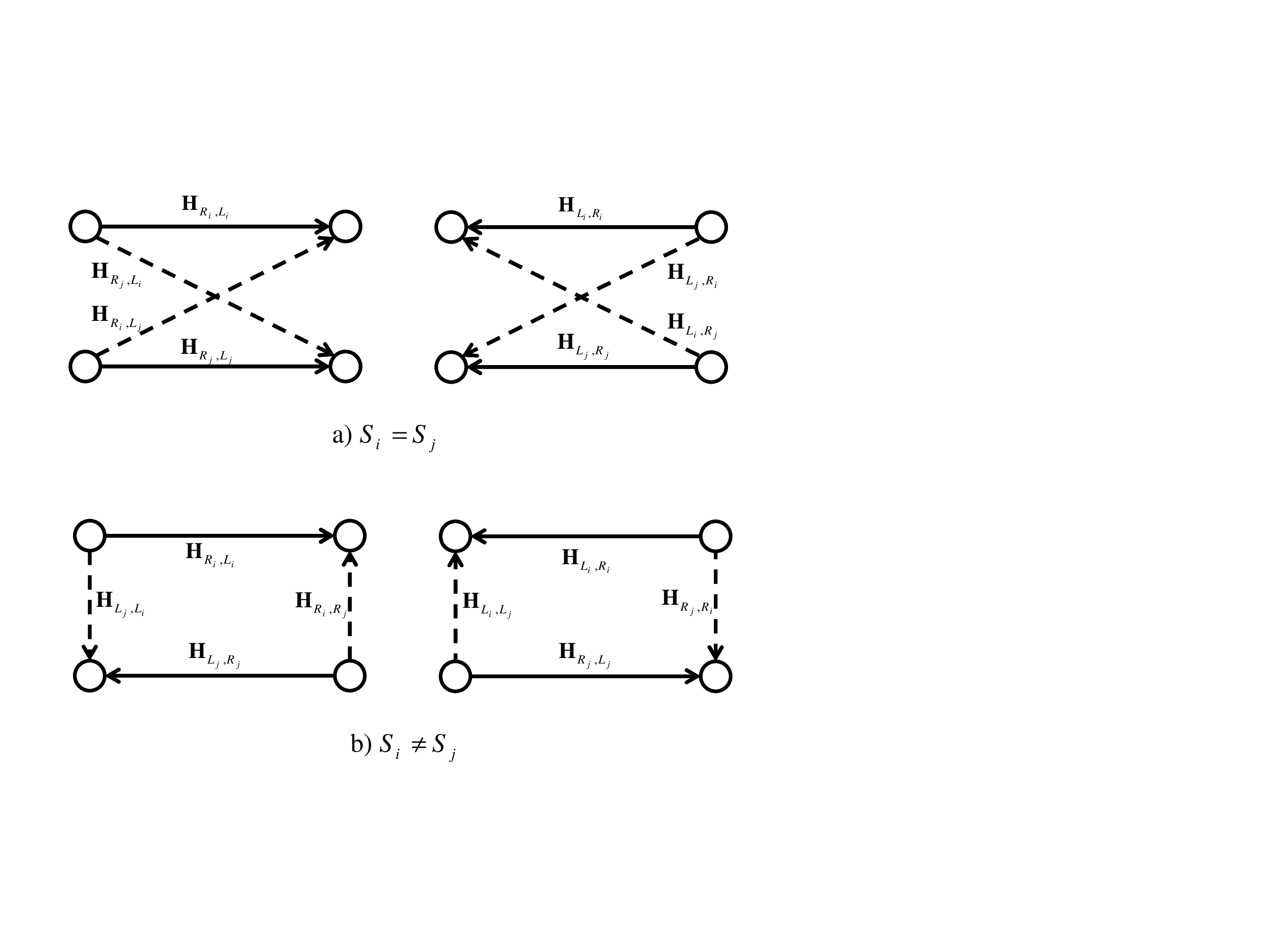}\protect\protect\protect\caption{\label{fig:3}Illustration of the definition of the MIMO channels.}
\end{figure}

\subsection{Signal Model}

Referring to Fig. \ref{fig:3} for an illustration of the main definitions,
the signal $\mathbf{y}_{R_{k}}\in\mathbb{C}^{N_{R}\times1}$ received
by the $k$th receiver when active in the $L_{k}\rightarrow R_{k}$
direction is given by
\begin{align}
\mathbf{y}_{R_{k}} & =\mathbf{H}_{R_{k},L_{k}}\mathbf{x}_{L_{k}}+\sum_{j\in\mathcal{K}\setminus\left\{ k\right\} }\left(1\left[S_{j}=S_{k}\right]\mathbf{H}_{R_{k},L_{j}}\mathbf{x}_{L_{j}}\right.\label{eq:received-signal}\\
 & \left.+1\left[S_{j}\neq S_{k}\right]\mathbf{H}_{R_{k},R_{j}}\mathbf{x}_{R_{j}}\right)+\mathbf{z}_{R_{k}},\nonumber
\end{align}
where the matrices $\mathbf{H}_{R_{k},L_{j}}\in\mathbb{C}^{N_{R}\times N_{L}}$
and $\mathbf{H}_{R_{k},R_{j}}\in\mathbb{C}^{N_{R}\times N_{R}}$ represent
the channel responses from the node $L_{j}$ to $R_{k}$, and from
the node $R_{j}$ to $R_{k}$, respectively; the vectors $\mathbf{x}_{L_{j}}$
and $\mathbf{x}_{R_{j}}$ are the signals transmitted by the nodes
$L_{j}$ and $R_{j}$, respectively; and the vector $\mathbf{z}_{R_{k}}$
is the additive noise, which is distributed as $\mathbf{z}_{R_{k}}\sim\mathcal{CN}(\mathbf{0},\sigma^{2}\mathbf{I})$
and is independent across the link index $k$. The first term in (\ref{eq:received-signal})
is the signal received from the desired transmitter while the second
term represents the interference from all the other links that have
the same spin as link $k$ and the third accounts for the interference
caused by links that have opposite spin. Similarly, the signal $\mathbf{y}_{L_{k}}\in\mathbb{C}^{N_{L}\times1}$
received by the $k$th receiver when active in the $R_{k}\rightarrow L_{k}$
direction can be written as
\begin{align}
\mathbf{y}_{L_{k}} & =\mathbf{H}_{L_{k},R_{k}}\mathbf{x}_{R_{k}}+\sum_{j\in\mathcal{K}\setminus\left\{ k\right\} }\left(1\left[S_{j}=S_{k}\right]\mathbf{H}_{L_{k},R_{j}}\mathbf{x}_{R_{j}}\right.\label{eq:received-signal-1}\\
 & \left.+1\left[S_{j}\neq S_{k}\right]\mathbf{H}_{L_{k},L_{j}}\mathbf{x}_{L_{j}}\right)+\mathbf{z}_{L_{k}},\nonumber
\end{align}
with analogous definitions.

The channel matrix $\mathbf{H}_{R_{k},L_{j}}$, and similarly for
the other pairs of nodes can be written as $\mathbf{H}_{R_{k},L_{j}}=\alpha_{R_{k},L_{j}}\tilde{\mathbf{H}}_{R_{k},L_{j}}$,
where the parameter $\alpha_{R_{k},L_{j}}$ models path-loss and shadowing
over the link $R_{k}-L_{j}$ in both directions $R_{k}\rightarrow L_{j}$
and $L_{j}\rightarrow R_{k}$ and is defined as
\begin{equation}
\alpha_{R_{k},L_{j}}=\sqrt{\left(\frac{D_{\textrm{ref}}}{D_{R_{k},L_{j}}}\right)^{\eta}\cdot10^{\xi_{R_{k},L_{j}}/10}},\label{eq:pathLoss_shadowing}
\end{equation}
where $D_{\textrm{ref}}$ is a reference distance, $D_{R_{k},L_{j}}$
is the distance between node $R_{k}$ and node $L_{j}$, $\eta$ is
the path loss exponent and $\xi_{R_{k},L_{j}}$ is the log-normal
shadowing component, which is distributed as $\xi_{R_{k},L_{j}}\sim\mathcal{N}(\mathbf{0},\nu^{2}\mathbf{I})$
and is independent for different pairs of nodes; and $\tilde{\mathbf{H}}_{R_{k},L_{j}}$
is an $N_{R}\times N_{L}$ channel matrix that accounts for the effect
of fast fading on the channel $L_{j}\rightarrow R_{k}$ and is assumed
to have i.i.d. $\mathcal{CN}(\mathbf{0},1)$ entries. The channel
matrices may or may not be reciprocal, so that, in general we have
the inequality $\tilde{\mathbf{H}}_{R_{k},L_{j}}\neq\tilde{\mathbf{H}}_{L_{j},R_{k}}^{H}$.

The channel matrices are assumed to be constant for at least a frame.
The analysis below will be limited to any period including frames
in which the channel matrices remain approximately constant.

When transmitting, nodes $L_{k}$ and $R_{k}$, first perform channel
coding, producing codeword vectors $\mathbf{s}_{L_{k}}\in\mathbb{C}^{d_{k}\times1}$
and $\mathbf{s}_{R_{k}}\in\mathbb{C}^{d_{k}\times1}$, which are distributed
as $\mathbf{s}_{L_{k}}\sim\mathcal{CN}(\mathbf{0},P_{k}/d_{k}\mathbf{I})$
and $\mathbf{s}_{R_{k}}\sim\mathcal{CN}(\mathbf{0},P_{k}/d_{k}\mathbf{I})$,
respectively, where parameter $d_{k}\leq\min\left(N_{R},N_{L}\right)$
represents the number of data streams exchanged on the two-way link
and $P_{k}$ is the received average signal-to-noise ratios, (SNRs),
excluding shadowing effects, at the reference distance $D_{\textrm{ref}}$.
Note that we assume that the power and the number of data streams
is the same in both directions for each link, although generalizations
are straightforward. We also assume for simplicity that equal power
allocation is performed across all data streams. After channel coding,
node $L_{k}$ performs linear precoding with a truncated unitary precoding
matrix $\mathbf{V}_{L_{k}}\in\mathbb{U}^{N_{L}\times d_{k}}$ yielding
the transmitted baseband signal
\begin{equation}
\mathbf{x}_{L_{k}}=\mathbf{V}_{L_{k}}\mathbf{s}_{L_{k}}\label{eq:precoding}
\end{equation}
and similarly node $R_{k}$ performs linear precoding with a truncated
unitary precoding matrix $\mathbf{V}_{R_{k}}\in\mathbb{U}^{N_{R}\times d_{k}}$
yielding the transmitted baseband signal
\begin{equation}
\mathbf{x}_{R_{k}}=\mathbf{V}_{R_{k}}\mathbf{s}_{R_{k}}
\end{equation}
On receiving the baseband signal $\mathbf{y}_{R_{k}}$, node $R_{k}$
performs linear equalization with a truncated unitary equalization
matrix $\mathbf{U}_{R_{k}}\in\mathbb{U}^{N_{R}\times d_{k}}$ yielding
the output signal
\begin{equation}
\hat{\mathbf{s}}_{L_{k}}=\mathbf{U}_{R_{k}}^{H}\mathbf{y}_{R_{k}},\label{eq:decoding}
\end{equation}
and similarly, on receiving the baseband signal $\mathbf{y}_{L_{k}}$,
node $L_{k}$ performs linear equalization with a truncated unitary
equalization matrix $\mathbf{U}_{L_{k}}\in\mathbb{U}^{N_{L}\times d_{k}}$
yielding the output signal
\begin{equation}
\hat{\mathbf{s}}_{R_{k}}=\mathbf{U}_{L_{k}}^{H}\mathbf{y}_{L_{k}}.
\end{equation}

The covariance matrix of the interference signals that affects the
equalized signal $\hat{\mathbf{s}}_{R_{k}}$, is given as
\begin{eqnarray}
\mathbf{Q}_{R_{k}}^{\textrm{RX}} & \negmedspace\negmedspace\negmedspace\negmedspace=\negmedspace\negmedspace\negmedspace & \mathbf{U}_{R_{k}}^{H}\negmedspace\negmedspace\left(\negmedspace\underset{j\in\mathcal{K}\setminus\{k\}}{\sum\negmedspace\negmedspace}\negmedspace\negmedspace\frac{P_{j}}{d_{j}}\negmedspace\left(\negmedspace1\negmedspace\left[S_{j}\negmedspace=\negmedspace S_{k}\right]\mathbf{H}_{R_{k},L_{j}}\mathbf{V}_{L_{j}}\mathbf{V}_{L_{j}}^{H}\mathbf{H}_{R_{k},L_{j}}^{H}\right.\right.\nonumber \\
 &  & \negmedspace\negmedspace\negmedspace\negmedspace\left.\left.+1\left[S_{j}\neq S_{k}\right]\mathbf{H}_{R_{k},R_{j}}\mathbf{V}_{R_{j}}\mathbf{V}_{R_{j}}^{H}\mathbf{H}_{R_{k},R_{j}}^{H}\negmedspace\right)\negmedspace\right)\negmedspace\mathbf{U}_{R_{k}}\negmedspace,\negmedspace\negmedspace\negmedspace\negmedspace\label{eq:8}
\end{eqnarray}
while for the equalized signal $\hat{\mathbf{s}}_{L_{k}}$ we have
\begin{eqnarray}
\mathbf{Q}_{L_{k}}^{\textrm{RX}} & \negmedspace\negmedspace\negmedspace\negmedspace=\negmedspace\negmedspace\negmedspace & \mathbf{U}_{L_{k}}^{H}\negmedspace\negmedspace\left(\negmedspace\underset{j\in\mathcal{K}\setminus\{k\}}{\sum\negmedspace\negmedspace}\negmedspace\negmedspace\frac{P_{j}}{d_{j}}\negmedspace\left(\negmedspace1\negmedspace\left[S_{j}\negmedspace=\negmedspace S_{k}\right]\mathbf{H}_{L_{k},R_{j}}\mathbf{V}_{R_{j}}\mathbf{V}_{R_{j}}^{H}\mathbf{H}_{L_{k},R_{j}}^{H}\right.\right.\nonumber \\
 &  & \negmedspace\negmedspace\negmedspace\negmedspace\left.\left.+1\left[S_{j}\neq S_{k}\right]\mathbf{H}_{L_{k},L_{j}}\mathbf{V}_{L_{j}}\mathbf{V}_{L_{j}}^{H}\mathbf{H}_{L_{k},L_{j}}^{H}\negmedspace\right)\negmedspace\right)\negmedspace\mathbf{U}_{L_{k}}\negmedspace.\negmedspace\negmedspace\negmedspace\negmedspace\label{eq:9}
\end{eqnarray}
It is also useful to define the covariance matrix of the overall interference
that is caused by node $R_{k}$ when transmitting as
\begin{eqnarray}
\mathbf{Q}_{R_{k}}^{\textrm{TX}} & \negmedspace\negmedspace\negmedspace\negmedspace=\negmedspace\negmedspace\negmedspace & \mathbf{V}_{R_{k}}^{H}\negmedspace\negmedspace\left(\!\underset{j\in\mathcal{K}\setminus\{k\}}{\sum\negmedspace\negmedspace}\negmedspace\negmedspace\frac{P_{j}}{d_{j}}\negmedspace\left(\negmedspace1\negmedspace\left[S_{j}\negmedspace=\negmedspace S_{k}\right]\mathbf{H}_{L_{j},R_{k}}^{H}\mathbf{U}_{L_{j}}\mathbf{U}_{L_{j}}^{H}\mathbf{H}_{L_{j},R_{k}}\right.\right.\nonumber \\
 &  & \negmedspace\negmedspace\negmedspace\negmedspace\left.\left.+1\left[S_{j}\neq S_{k}\right]\mathbf{H}_{R_{j},R_{k}}^{H}\mathbf{U}_{R_{j}}\mathbf{U}_{R_{j}}^{H}\mathbf{H}_{R_{j},R_{k}}\negmedspace\right)\negmedspace\right)\negmedspace\mathbf{V}_{R_{k}}\negmedspace,\negmedspace\negmedspace\negmedspace\negmedspace\label{eq:10}
\end{eqnarray}
while the covariance matrix of the overall interference that is caused
by node $L_{k}$ when transmitting is given as
\begin{eqnarray}
\mathbf{Q}_{L_{k}}^{\textrm{TX}} & \negmedspace\negmedspace\negmedspace\negmedspace=\negmedspace\negmedspace\negmedspace & \mathbf{V}_{L_{k}}^{H}\negmedspace\negmedspace\left(\negmedspace\underset{j\in\mathcal{K}\setminus\{k\}}{\sum\negmedspace\negmedspace}\negmedspace\negmedspace\frac{P_{j}}{d_{j}}\negmedspace\left(\negmedspace1\negmedspace\left[S_{j}\negmedspace=\negmedspace S_{k}\right]\mathbf{H}_{R_{j},L_{k}}^{H}\mathbf{U}_{R_{j}}\mathbf{U}_{R_{j}}^{H}\mathbf{H}_{R_{j},L_{k}}\right.\right.\nonumber \\
 &  & \negmedspace\negmedspace\negmedspace\negmedspace\left.\left.+1\left[S_{j}\neq S_{k}\right]\mathbf{H}_{L_{j},L_{k}}^{H}\mathbf{U}_{L_{j}}\mathbf{U}_{L_{j}}^{H}\mathbf{H}_{L_{j},L_{k}}\negmedspace\right)\negmedspace\right)\negmedspace\mathbf{V}_{L_{k}}\negmedspace.\negmedspace\negmedspace\negmedspace\negmedspace\label{eq:11}
\end{eqnarray}
Note that the defined interference covariance matrices depend on the
interference spin variables $S_{k}$, $k\in{\cal K}$, and on the
linear precoding and equalization matrices.

\section{Bi-Directional Scheduling and Interference Alignment Optimization\label{sec:Scheduling approaches}}

In this section, we propose algorithms that perform the optimization
of the interference spins vector $\mathbf{S}=\left[S_{1},...,S_{K}\right]$
and of the linear transceivers $\mathbf{U}=[\left\{ \mathbf{U}_{R_{k}}\right\} _{k=1}^{K},\left\{ \mathbf{U}_{L_{k}}\right\} _{k=1}^{K}]$
and $\mathbf{V}=[\left\{ \mathbf{V}_{R_{k}}\right\} _{k=1}^{K},\left\{ \mathbf{V}_{L_{k}}\right\} _{k=1}^{K}]$.
Following \cite{Jafar3}, in order to approximate the interference
alignment conditions, we adopt as the optimization criterion the interference
power leakage. Note that \cite{Jafar3} only considers one-way communication
links and hence the optimization therein is limited to the linear
transceivers. Specifically, we formulate the problem of interest as
the minimization of the total received interference leakage power
as
\begin{eqnarray}
\underset{\mathbf{S},\mathbf{U},\mathbf{V}}{\textrm{minimize}} & \sum_{k\in\mathcal{K}}I_{R_{k}}^{\textrm{RX}}+I_{L_{k}}^{\textrm{RX}},\label{eq:12}
\end{eqnarray}
or, equivalently, as
\begin{eqnarray}
\underset{\mathbf{S},\mathbf{U},\mathbf{V}}{\textrm{minimize}} & \sum_{k\in\mathcal{K}}I_{R_{k}}^{\textrm{TX}}+I_{L_{k}}^{\textrm{TX}},\label{eq:12-1}
\end{eqnarray}
where we have the implicit constraints that $\mathbf{S}\in\{0,1\}^{K}$
hold and that the matrices $\mathbf{U}$ and $\mathbf{V}$ be truncated
unitary with the mentioned dimensions; moreover, the quantities
\begin{eqnarray}
I_{R_{k}}^{\textrm{RX}} & = & \textrm{tr}\left(\mathbf{Q}_{R_{k}}^{\textrm{RX}}\right)\textrm{ and }I_{L_{k}}^{\textrm{RX}}=\textrm{tr}\left(\mathbf{Q}_{L_{k}}^{\textrm{RX}}\right)\label{eq:13}
\end{eqnarray}
measure the received interference power for link $k$ in the two directions,
while
\begin{eqnarray}
I_{R_{k}}^{\textrm{TX}} & = & \textrm{tr}\left(\mathbf{Q}_{R_{k}}^{\textrm{TX}}\right)\textrm{ and }I_{L_{k}}^{\textrm{TX}}=\textrm{tr}\left(\mathbf{Q}_{L_{k}}^{\textrm{TX}}\right)\label{eq:15}
\end{eqnarray}
measure the interference power caused by the transmitters of link
$k$ in the two directions.

We observe that it is possible to extend the proposed approach to
other performance criteria including the max-SINR method of \cite{Jafar3},
but this will not be further pursued here. Instead, the focus is on
approximating the interference alignment conditions as per the interference
leakage minimization (ILM) scheme in \cite{Jafar3}. We start by discussing
an algorithm that performs the separate optimization of the interference
spins and of the linear transceivers in Sec. \ref{sub:Sep_opt} and
then we introduce an algorithm that carries out the joint optimization
of interference spins and linear transceivers in Sec. \ref{sub:Joint-Optimization}.

\subsection{Separate Optimization \label{sub:Sep_opt}}

A first simple solution is that of first determining the spin variables
$\mathbf{S}$ and then perform linear transceiver optimization using
the ILM algorithm as in \cite{Jafar3}.

\textbf{Spin Optimization:} In order to optimize the spin variables,
we propose to solve the problem of (\ref{eq:12}), or equivalently
(\ref{eq:12-1}), by setting all precoding and equalization matrices
equal to the identity matrix in (\ref{eq:8})-(\ref{eq:11}). This
optimization can either be carried out by exhaustive search in the
space $\{0,1\}^{K}$ or by using the algorithm proposed in \cite{Jimmy}.
Moreover, in order to minimize the overhead associated to the selection
of the vector $\mathbf{S}$, one can perform the optimization of $\mathbf{S}$
based only on long-term CSI, namely path loss and log-normal shadowing.
This enables the spins to be updated only at the time scale of the
long-term fading variability. In this case, problem (\ref{eq:12}),
or (\ref{eq:12-1}), is tackled by setting all matrices $\mathbf{U}$
and $\mathbf{V}$ to the identity matrix and by averaging the interference
powers over the fast fading variables. This leads to the problem
\begin{eqnarray}
\underset{\mathbf{S}}{\textrm{min}\negmedspace\negmedspace}\negmedspace\negmedspace & \underset{k\in\mathcal{K}}{\sum\negmedspace\negmedspace\negmedspace\negmedspace} & \negmedspace\negmedspace\negmedspace\negmedspace\left(\negmedspace\underset{j\in\mathcal{K}\setminus\{k\}}{\sum\negmedspace\negmedspace}\negmedspace\negmedspace\left(\negmedspace1\left[S_{j}\negmedspace=\negmedspace S_{k}\right]\left(\negmedspace\frac{D_{\textrm{ref}}}{D_{R_{k},L_{j}}}\negmedspace\right)^{\eta}\negmedspace\negmedspace\cdot\negmedspace10^{\frac{\xi_{R_{k},L_{j}}}{10}}N_{L}N_{R}\right.\right.\nonumber \\
\negmedspace\negmedspace\negmedspace\negmedspace & \negmedspace\negmedspace\negmedspace\negmedspace & \negmedspace\negmedspace\negmedspace\negmedspace+1\left[S_{j}\negmedspace\neq\negmedspace S_{k}\right]\negmedspace\left(\negmedspace\frac{D_{\textrm{ref}}}{D_{R_{k},R_{j}}}\negmedspace\right)^{\negmedspace\eta}\negmedspace\negmedspace\cdot\negmedspace10^{\frac{\xi_{R_{k},R_{j}}}{10}}N_{R}^{2}\nonumber \\
\negmedspace\negmedspace\negmedspace\negmedspace & \negmedspace\negmedspace\negmedspace\negmedspace & \negmedspace\negmedspace\negmedspace\negmedspace1\left[S_{j}\negmedspace=\negmedspace S_{k}\right]\negmedspace\left(\negmedspace\frac{D_{\textrm{ref}}}{D_{L_{k},R_{j}}}\negmedspace\right)^{\negmedspace\eta}\negmedspace\negmedspace\cdot\negmedspace10^{\frac{\xi_{L_{k},R_{j}}}{10}}N_{L}N_{R}\nonumber \\
\negmedspace\negmedspace\negmedspace\negmedspace & \negmedspace\negmedspace\negmedspace\negmedspace & \negmedspace\negmedspace\negmedspace\negmedspace\left.\left.+1\left[S_{j}\negmedspace\neq\negmedspace S_{k}\right]\negmedspace\left(\negmedspace\frac{D_{\textrm{ref}}}{D_{L_{k},L_{j}}}\negmedspace\right)^{\negmedspace\eta}\negmedspace\negmedspace\cdot\negmedspace10^{\frac{\xi_{L_{k},L_{j}}}{10}}N_{L}^{2}\right)\right),\negmedspace\negmedspace\negmedspace\negmedspace\label{eq:16}
\end{eqnarray}
which can be tackled as explained above.

\begin{figure}
\centering\includegraphics[clip,scale=0.43]{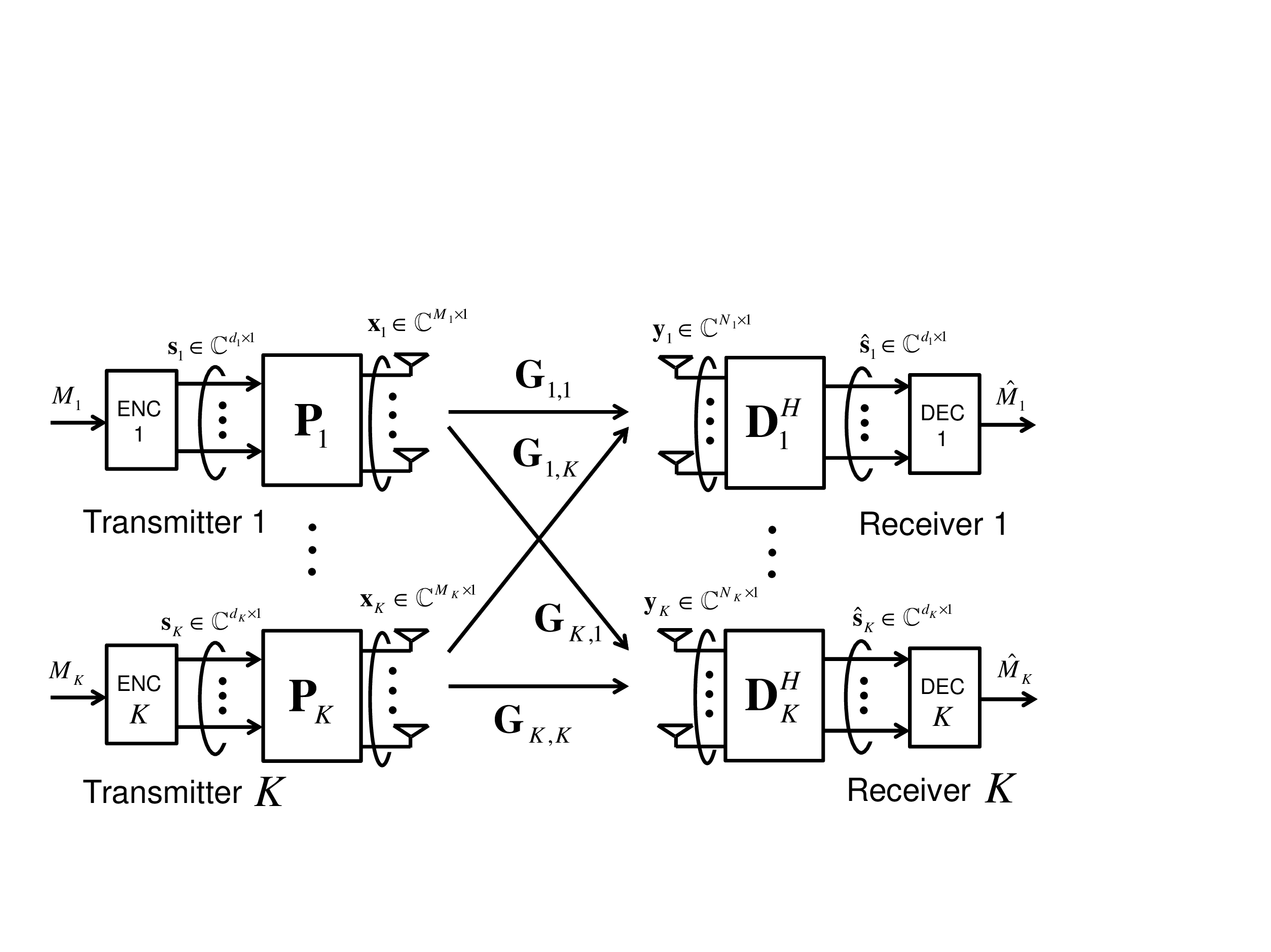}

\protect\protect\protect\caption{\label{fig:4}Illustration of a $K$-user MIMO interfering channel.}
\end{figure}

\textbf{Interference Leakage Minimization (ILM): }After finding the
optimized spin vector $\mathbf{S}=\left[S_{1},S_{2},...,S_{K}\right]$
as explained above, the linear transceiver matrices $\mathbf{V}$
and $\mathbf{U}$ can be calculated using the ILM method of \cite[Alg. 1]{Jafar3}
using full CSI. The ILM algorithm aims at minimizing the interference
power over a $K$-user MIMO interfering channel with generic channel
matrices $\left\{ \mathbf{G}_{k,j}\in\mathbb{C}^{N_{k}\times M_{j}}\right\} _{k,j=1}^{K}$
over the precoding matrices $\left\{ \mathbf{P}_{k}\in\mathbb{U}^{M_{k}\times d_{k}}\right\} _{k=1}^{K}$
and decoding matrices $\left\{ \mathbf{D}_{k}\in\mathbb{U}^{N_{k}\times d_{k}}\right\} _{k=1}^{K}$
as illustrated in Fig. \ref{fig:4}. The algorithm runs over a given
number $N_{\textrm{ILM}}$ of iterations and is based on the calculation,
which can be implemented in a decentralized fashion, of the leading
eigenvalues of given covariance matrices \cite{Jafar3}. We denote
the outcome of the ILM algorithm as
\begin{align}
\negmedspace\negmedspace\left[\left\{ \mathbf{P}_{j}\right\} _{j=1}^{K},\left\{ \mathbf{D}_{j}\right\} _{j=1}^{K}\right] & \negmedspace\negmedspace=\negmedspace\textrm{ILM}\left(\left\{ \mathbf{G}_{k,j}\right\} _{k,j=1}^{K},\left\{ d_{j}\right\} _{j=1}^{K}\right.,\nonumber \\
\negmedspace\negmedspace\negmedspace\negmedspace & \left.N_{\textrm{ILM}},\left\{ \mathbf{P}_{j}\right\} _{j=1}^{K},\left\{ \mathbf{D}_{j}\right\} _{j=1}^{K}\right)\negmedspace,\negmedspace
\end{align}
where we do not make explicit the dependence on the initial conditions
for simplicity of notation. We recall the ILM algorithm for convenience
in Table Algorithm \ref{alg:Algorithm1}.
\begin{algorithm}
\protect\caption{\label{alg:Algorithm1}Interference Leakage Minimization \cite{Jafar3}}

1: Start with initial precoding matrices $\mathbf{P}_{j}$ with $\mathbf{P}_{j}^{H}\mathbf{P}_{j}=\mathbf{I}$
for all $j\in{\cal K}$ .

2: Begin iteration with setting $i=1$

3: Compute interference covariance matrix at the receivers $k\in{\cal K}$:
\begin{eqnarray*}
\mathbf{Q}_{k}^{\textrm{RX}} & = & \sum_{j\in\mathcal{K}\setminus\{k\}}\mathbf{G}_{k,j}\!\mathbf{P}_{j}\!\mathbf{P}_{j}^{H}\!\mathbf{G}_{k,j}^{H}\!
\end{eqnarray*}
4: Compute the linear equalization matrix at each receiver $k\in{\cal K}$:
\[
\mathbf{D}_{k}=\nu_{\ce{min}}^{d_{k}}\left(\mathbf{Q}_{k}^{\textrm{RX}}\right)
\]

5: Compute interference covariance matrix for each transmitter $j$
with $j\in{\cal K}$:
\begin{eqnarray*}
\mathbf{Q}_{j}^{\textrm{TX}} & = & \sum_{k\in\mathcal{K}\setminus\{j\}}\mathbf{G}_{k,j}^{H}\!\mathbf{D}_{k}\!\mathbf{D}_{k}^{H}\!\mathbf{G}_{k,j}\!
\end{eqnarray*}
6: Compute the linear precoding matrix for each transmitter $j$ with
$j\in{\cal K}$:
\[
\mathbf{P}_{j}=\nu_{\ce{min}}^{d_{j}}\left(\mathbf{Q}_{j}^{\textrm{TX}}\right)
\]

7: If $i>N_{\textrm{ILM}}$, exit; otherwise $i=i+1$ and go back
to 3.
\end{algorithm}

Based on the given vector $\mathbf{S}$, we apply ILM separately to
the even and odd slots. For the even slots, we assign the corresponding
channel responses matrices $\mathbf{H}_{R_{k},L_{j}}$, $\mathbf{H}_{R_{k},R_{j}}$,
$\mathbf{H}_{L_{k},R_{j}}$ and $\mathbf{H}_{L_{k},L_{j}}$ to $\mathbf{G}_{k,j}$
such that
\begin{align}
\mathbf{G}_{k,j} & =1\left[S_{j}=0\right]1\left[S_{j}=S_{k}\right]\mathbf{H}_{R_{k},L_{j}}\nonumber \\
 & +1\left[S_{j}=0\right]1\left[S_{j}\neq S_{k}\right]\mathbf{H}_{L_{k},L_{j}}\nonumber \\
 & +1\left[S_{j}=1\right]1\left[S_{j}=S_{k}\right]\mathbf{H}_{L_{k},R_{j}}\nonumber \\
 & +1\left[S_{j}=1\right]1\left[S_{j}\neq S_{k}\right]\mathbf{H}_{R_{k},R_{j}}\label{eq:18}
\end{align}
and
\begin{equation}
\mathbf{P}_{k}=1\left[S_{k}=0\right]\mathbf{V}_{L_{k}}+1\left[S_{k}=1\right]\mathbf{V}_{R_{k}}\label{eq:19}
\end{equation}
and
\begin{equation}
\mathbf{D}_{k}=1\left[S_{k}=1\right]\mathbf{U}_{L_{k}}+1\left[S_{k}=0\right]\mathbf{U}_{R_{k}}\label{eq:20}
\end{equation}
for all $j,k\in{\cal K}$. Instead, for the odd time slots, the assignment
above are obtained by substituting $\mathbf{S}$ with its entry-wise
complement $\bar{\mathbf{S}}$.

\subsection{Joint Optimization\label{sub:Joint-Optimization}}

In this section, we propose a technique that carries out the joint
optimization of interference spins and linear transceivers based on
full CSI. In order to avoid the exponential complexity of exhaustive
search, which scales as $2^{K}$, the proposed scheme explores a subset
of possible spin vectors guided by a simple local search criterion.
Specifically, at each step $i$, the algorithm explores a new spin
vector $\mathbf{S}^{(i)}$ based on the previous vector $\mathbf{S}^{(i-1)}$
by flipping the spins in $\mathbf{S}^{(i-1)}$ corresponding to the
minimum number of links, causing the largest transmitted interference
powers. More precisely, we define the set $\mathcal{S}{}^{(i-1)}$
to include all spin vectors explored up to iteration $(i-1)$ and
as $I_{\textrm{opt}}^{(i-1)}$ the minimum interference leakage power
observed up to iteration $(i-1)$ along with the corresponding spin
vector $\mathbf{S}_{\textrm{opt}}^{(i-1)}$. Moreover, we define as
$I_{R_{k}}^{\textrm{TX }(i-1)}$ and $I_{L_{k}}^{\textrm{TX }(i-1)}$
the interference powers (\ref{eq:15}) calculated using ILM as discussed
in the previous subsection for the current spin vector $\mathbf{S}^{(i-1)}$.
At each step $i$, we flip the spin variables of the vector $\mathbf{S}^{(i-1)}$
in the order of decreasing caused interference power $I_{R_{k}}^{\textrm{TX }(i-1)}+I_{L_{k}}^{\textrm{TX }(i-1)}$
until a new spin vector $\mathbf{S}^{(i)}\notin\mathcal{S}{}^{(i-1)}$
is obtained. The variables $\mathbf{S}^{(i)}$, $I_{R_{k}}^{\textrm{TX }(i)}$,
$I_{L_{k}}^{\textrm{TX }(i)}$, $I_{\textrm{opt}}^{(i)}$ and $\mathbf{S}_{\textrm{opt}}^{(i)}$
are updated accordingly based on ILM for the vector $\mathbf{S}^{(i)}$.
We perform a number $F$ of comparison steps with $F\leq2^{K}$.
\begin{rem}
\label{rem:RX_intf_criterion}At each step $i$, rather than using
the largest transmitted interference powers criterion, we could generate
a new vector $\mathbf{S}^{(i)}$ by flipping the spin of the link,
or links, with the largest \textit{received} interference powers,
namely $I_{R_{k}}^{\textrm{RX }(i-1)}+I_{L_{k}}^{\textrm{RX }(i-1)}$,
when using the previous spin vector $\mathbf{S}^{(i-1)}$. Extensive
numerical results, discussed in part in the next section, reveal that
the transmitted interference power criterion is generally to be preferred.
\end{rem}
\begin{algorithm}
\protect\protect\protect\caption{\label{alg:Algorithm2}Joint Spin and Linear Transceiver Optimization}

1: Initialize the vector $\mathbf{S}^{(0)}$and set $\mathbf{S}_{\textrm{opt}}^{(0)}=\mathbf{S}^{(0)}$,
$\mathcal{S}{}^{(-1)}=\emptyset$, and $I_{\textrm{opt}}^{(0)}$ to
a very large value.

2: Begin step $i=1$.

3: Run two instances of $\textrm{ILM }\left(\left\{ \mathbf{G}_{k,j}\right\} _{k,j=1}^{K},\left\{ d_{j}\right\} _{j=1}^{K},N_{\textrm{ILM}},\left\{ \mathbf{P}_{j}\right\} _{j=1}^{K},\left\{ \mathbf{D}_{j}\right\} _{j=1}^{K}\right)$
with (\ref{eq:18})-(\ref{eq:20}), one instance with $\mathbf{S}^{(i-1)}$
in lieu of $\mathbf{S}$ and one with $\bar{\mathbf{S}}^{(i-1)}$
in lieu of $\mathbf{S}$. Calculate the resulting total two-way transmitted
interference leakage, i.e., $I_{k}^{\textrm{TX }(i-1)}=I_{R_{k}}^{\textrm{TX }(i-1)}+I_{L_{k}}^{\textrm{TX }(i-1)}$,
and received interference leakage, i.e., $I_{k}^{\textrm{RX }(i-1)}=I_{R_{k}}^{\textrm{RX }(i-1)}+I_{L_{k}}^{\textrm{RX }(i-1)}$,
for all links $k\in\mathcal{K}$.

4: If $\sum_{k}I_{k}^{\textrm{TX }(i-1)}+I_{k}^{\textrm{RX }(i-1)}\leq I_{\textrm{opt}}^{(i-1)}$,
set $I_{\textrm{opt}}^{(i)}=I^{(i-1)}$ and $\mathbf{S}_{\textrm{opt}}^{(i)}=\mathbf{S}^{(i-1)}$;
otherwise, set $I_{\textrm{opt}}^{(i)}=I_{\textrm{opt}}^{(i-1)}$
and $\mathbf{S}_{\textrm{opt}}^{(i)}=\mathbf{S}_{\textrm{opt}}^{(i-1)}$.

5: Update the set of explored spin vectors $\mathcal{S}^{(i-1)}=\mathcal{S}^{(i-2)}\cup\{\mathbf{S}^{(i-1)}\}$.

6: Flip the smallest number of spin variables of the vector $\mathbf{S}^{(i-1)}$
in the order of decreasing caused interference power $I_{R_{k}}^{\textrm{TX }(i-1)}+I_{L_{k}}^{\textrm{TX }(i-1)}$
until a new spin vector $\mathbf{S}^{(i)}\notin\mathcal{S}{}^{(i-1)}$
is obtained.

7: If $i>F$, stop, otherwise set $i=i+1$ and go back to 4.
\end{algorithm}

\section{Numerical Results}

In this section, we evaluate the performance of the considered techniques
via numerical results. We consider a $100m\times100m$ area in where
all the links are located randomly. We set $\eta=3$ and $\xi=5dB$
for all links. There are $K_{1}$ shorter-distance links with a transmitter-receiver
distance of $D_{1}=25m$ and $K_{2}$ longer-distance links with distance
of $D_{2}=50m$, so that $K=K_{1}+K_{2}$. The shorter-distance links
may represent D2D connections and the longer-distance links may represent
WLAN channels between a device and an access point. In keeping with
this interpretation, it is also assumed that both nodes of a shorter-distance
link have the same average SNRs at the reference distance $D_{\textrm{ref}}=D_{1}$
of $P_{L}=P_{R}=10$dB, while for the longer-distance links we have
$P_{L}=30$dB and $P_{R}=50$dB. Note that the corresponding average
SNRs at the distance $D_{2}$ are $P_{L}(D_{\textrm{ref}}/D_{2})^{\eta}=21$dB
and $P_{R}(D_{\textrm{ref}}/D_{2})^{\eta}=41$dB. The number of antennas
and the number of data streams are the same for all the nodes, namely
$N_{L}=N_{R}=4$ and $d=2$, respectively. The number of ILM iterations
is $N_{\textrm{ILM}}=20$.

\begin{figure}
\centering\includegraphics[clip,scale=0.5]{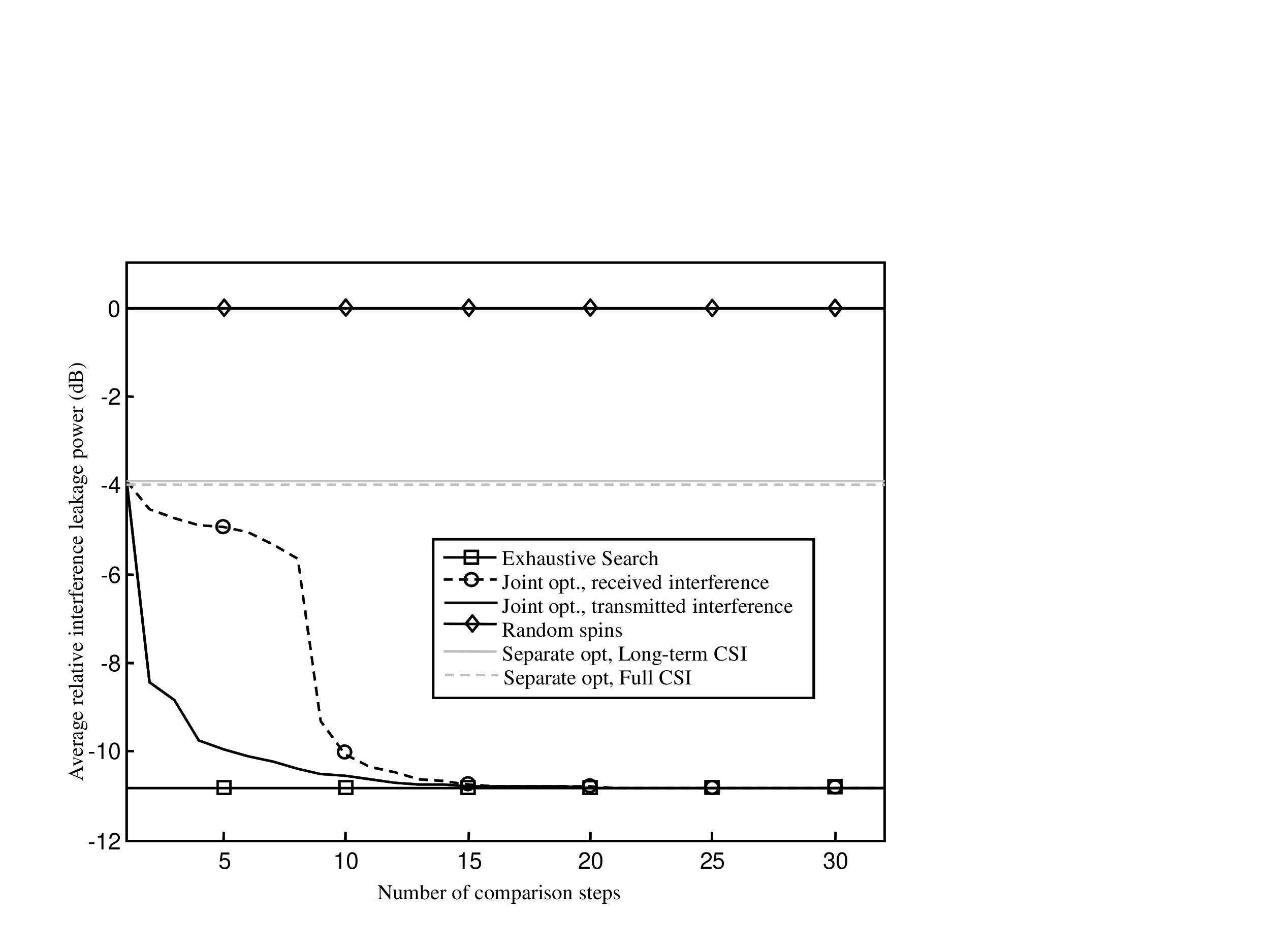}\protect\protect\protect\caption{\label{fig:5}Average relative interference leakage power (dB) versus
the number of comparison steps for the proposed approach and for reference
schemes ($K_{1}=3$, $K_{2}=2$, $N_{L}=N_{R}=4$, $d=2$, and $N_{\textrm{ILM}}=20$).}
\end{figure}

Fig. \ref{fig:5} illustrates the average interference leakage power
versus the number of comparison steps of the proposed Algorithm \ref{alg:Algorithm2},
with $K_{1}=3$ and $K_{2}=2$. For comparison, we show the performance
of exhaustive search, of ILM run on a randomly selected spin vector
and of the separate optimization schemes discussed in Sec. \ref{sub:Sep_opt}.
For Algorithm \ref{alg:Algorithm2}, the spin initialization is done
using the result of separate optimization using long-term CSI, i.e.,
by solving problem (\ref{eq:16}). The average interference leakage
is normalized to the average interference leakage power obtained with
random spins. The performance of the proposed techniques is seen to
be very close to that of exhaustive search as long as around 10 comparison
steps are performed. This is substantially smaller than the required
number of comparison steps in exhaustive search, i.e., $2^{5}=32$.
This advantage is less pronounced if one uses instead the received
interference power criterion, as discussed in Remark \ref{rem:RX_intf_criterion}.
Moreover, the interference leakage power can be significantly reduced
with respect to the considered baseline techniques, e.g., by more
than 10dB over a random spin selection. This demonstrates that the
interference alignment capabilities of the system are enhanced by
the adoption of joint optimization. Finally, we observe that separate
optimization based on long-term CSI perform very similarly to separate
optimization based on full CSI.

\begin{figure}
\centering\includegraphics[clip,scale=0.5]{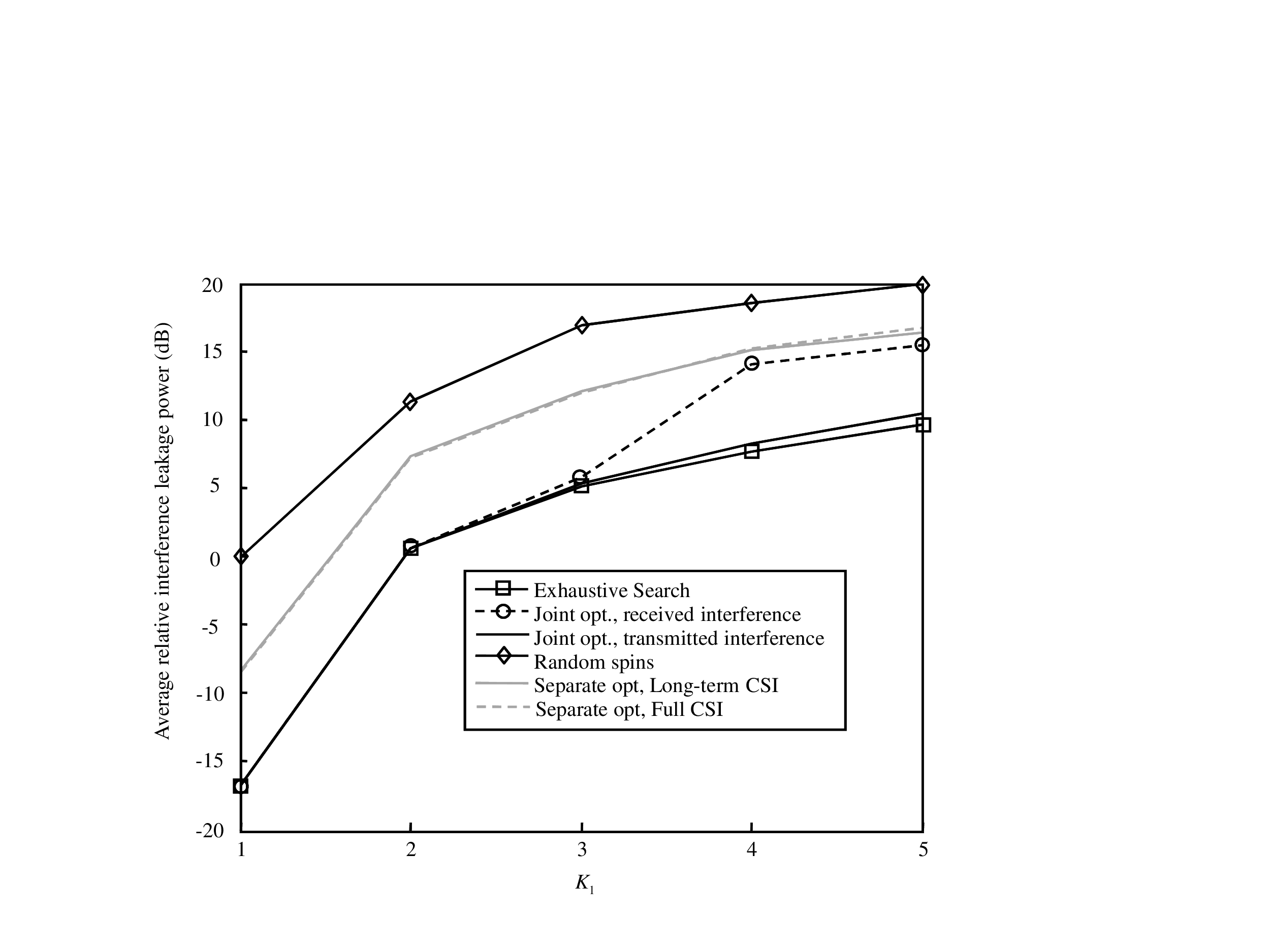}\protect\protect\protect\caption{\label{fig:6}Average relative interference leakage power (dB) versus
the number of shorter-distance links $K_{1}$ ($K_{2}=2$, $N_{L}=N_{R}=4$,
$d=2$, and $N_{\textrm{ILM}}=20$).}
\end{figure}

Fig. \ref{fig:6} shows the average interference leakage versus the
number of shorter-distance links with $K_{2}=2$ and the number of
comparison steps set to 10 for the proposed algorithm. The interference
leakage power is normalized to the average interference leakage power
obtained with random spins. While the interference leakage power increases
with $K_{1}$, the relative performance of all schemes remains fairly
constant, except for the algorithm based on the received interference
power (see Remark \ref{rem:RX_intf_criterion}), which, as seen above,
tends to require a larger number of comparison steps when $K_{1}$
is large enough.

\begin{figure}
\centering\includegraphics[clip,scale=0.5]{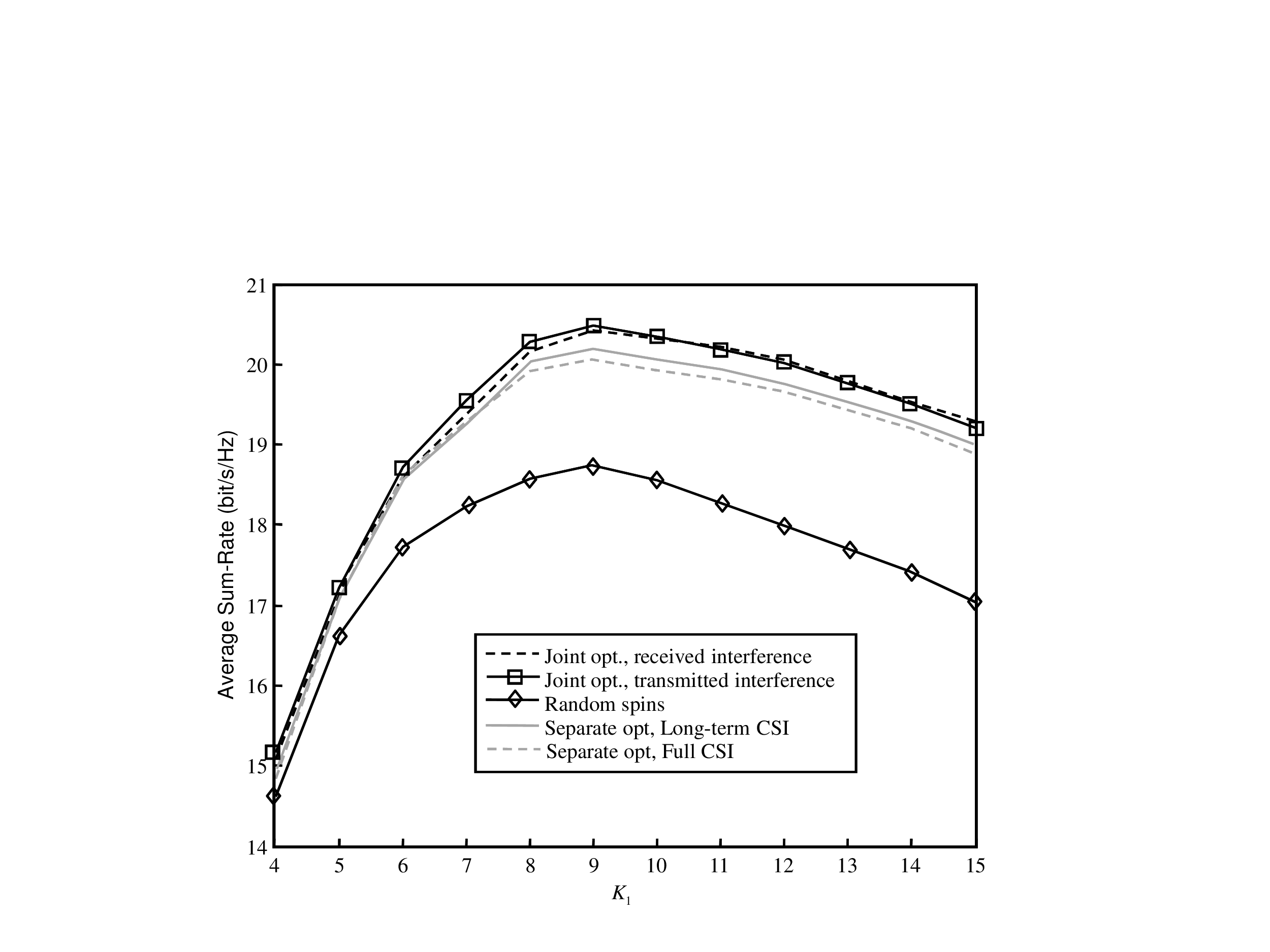}\protect\protect\protect\caption{\label{fig:7}Average sum-rate versus the number of links $K_{1}$
($K_{2}=0$, $N_{L}=N_{R}=4$, $d=2$, and $N_{\textrm{ILM}}=20$).}
\end{figure}

Fig. \ref{fig:7} shows the average sum-rate, which is calculated
using standard capacity formulas (see, e.g., \cite{Jafar3}), versus
the number of links $K_{1}$ with $K_{2}=0$. The relevant gains of
optimizing the spin vector are confirmed to hold also in terms of
achievable two-way rate. Separate optimization is, however, shown
to reap most of this advantage, even when using only long-term CSI.
Finally, we note that increasing the number of links $K_{1}$ is at
first beneficial as long as interference is manageable, but it leads
to a decrease of the rate as soon as the number of links is large
enough (here, larger than 9).

\section{Concluding Remarks}

In this paper, we have investigated a wireless network with MIMO two-way
interfering links that operate using TDD with fixed switching times.
We have proposed to optimize the bi-directional schedule, i.e., the
order of the transmission directions, along with the linear transceivers,
with the aim of minimizing the interference leakage power. Numerical
results show that the performance of the proposed technique in terms
of interference leakage is very close to that of a scheme based on
exhaustive search, but at a substantially lower complexity, and that
significant gains can be obtained as compared to a random selection
of the interference spins. Interestingly, it is also seen that the
interference leakage gains do not necessarily translate into sum-rate
gains, for which separate optimization with long-term CSI appear to
be sufficient to reap most of the performance advantage. Interesting
future work includes the investigation of a distributed version of
the proposed scheme.

\end{document}